\tolerance=10000
\input phyzzx

\REF\VafaWitten{C. Vafa and E. Witten,  {\it Nucl. Phys. } {\bf B431} (1994)
77.}
\REF\VariousDim{L. Baulieu, H. Kanno and I. Singer, {\it Commun.
Math. Phys.} {\bf 194} (1998) 149; L. Baulieu, A. Losev and N.
Nekrasov, {\it Nucl. Phys.} {\bf B522} (1998) 82; L. Baulieu and
P. West, {\it Phys. Lett.} {\bf B436} (1998) 97-107; M. Blau and
G. Thompson, {\it Phys. Lett.} {\bf B415} (1997) 242; B.S.
Acharya, M. O'Loughlin and B. Spence, {\it Nucl. Phys.} {\bf B503}
(1997) 657; B.S. Acharya, J.M. Figueroa-O'Farrill, M. O'Loughlin
and B. Spence, {\it Nucl. Phys.} {\bf B514} (1998) 583; J.M.
Figueroa-O'Farrill, A. Imaanpur, J. McCarthy,
   {\it Phys. Lett.} {\bf B419} (1998) 167.}
\REF\WittenWeyl{E. Witten,  {\it Phys. Lett.} {\bf B206} (1988) 601.}
\REF\Fre{D. Anselmi and P. Fre,  {\it Nucl. Phys.} {\bf B392} (1993) 401.}
\REF\Laters{D. Anselmi, {\it Class. Quant. Grav.} {\bf 14} (1997)
2031-2047; M. Abe, {\it Mod. Phys. Lett} {\bf A12} (1997) 381-392;
P.L. Paul, {\it Topological Symmetries of Twisted N=2 Chiral
Supergravity in Ashtekar Formalism}, {\tt hep-th/9504144}. }
\REF\Baul{L. Baulieu and A. Tanzini, {\it JHEP} {\bf 03} (2002) 015.}
\REF\FvN{S. Ferrara and P. van Nieuwenhuizen, {\it Phys. Rev. Lett.} {\bf 37}
25 (1976) 1669.}
\REF\NisSez{H. Nishino and E. Sezgin, {\it Nucl. Phys.} {\bf B278} (1986) 353-379.}
\REF\TvN{U. Theis and P. van Nieuwenhuizen, {\it Class. Quant.
Grav.} {\bf 18} (2001) 5469-5486.}
\REF\Weinberg{S. Weinberg, {\it Gravitation and Cosmology}, John Wiley 1972,
Section 12.5.}
\REF\BaulTwo{L. Baulieu, M. Bellon and A. Tanzini, {\it Eight-Dimensional
Topological Gravity and its
Correspondence with Supergravity}, PAR-LPTHE 02-06, {\tt hep-th/0207020}.}
\REF\bill{B. Spence,{\it Topological Born-Infeld Actions and D-Branes},
{\tt hep-th/9907053}.}
\REF\vafaetal{R. Gopakumar and C. Vafa, {\it Adv. Theor. Math.
Phys.}, {\bf 3} (1999) 1415, {\tt hep-th/9811131}; J. Labastida and M.
Mari\t no,
{\it Journal of Knot Theory and Its Ramifications}, {\bf 11} 2 (2002) 173-197,
 {\tt math.QA/0104180}.}
\REF\dfhs{P. de Medeiros, J. Figueroa-O'Farrill, C. Hull and B.
Spence, {\it Conformal topological Yang-Mills theory and de Sitter
holography}, {\it JHEP} {\bf 08} (2002) 055, {\tt hep-th/0111190}.}


\Pubnum{ \vbox{  \hbox {QMUL-PH-02-18} \hbox{hep-th/0209115} }}
\pubtype{}
\date{September 2002}
\date{revised March 2003}
\titlepage
\title {\bf Four-dimensional topological Einstein-Maxwell gravity}
\author{P. de Medeiros and B. Spence}
\address{Physics Department,
\break
Queen Mary, University of London,
\break
Mile End Road, London E1 4NS, UK}

\vskip 2cm
 \centerline{\it In memory of Sonia Stanciu 1966-2002}

\abstract {The complete on-shell action of topological
Einstein-Maxwell gravity in four-dimensions is presented.
It is shown explicitly how this
theory for $SU(2)$ holonomy manifolds
arises from four-dimensional Euclidean $N=2$ supergravity.
The twisted local BRST symmetries and twisted local Lorentz
symmetries are given and the action and stress tensor are shown to
be BRST-exact. A set of BRST-invariant topological operators is
given. The local vector and antisymmetric tensor twisted supersymmetries
and their algebra are also found.}

\endpage


\chapter{Introduction}

The study of topological gauge theories is now a large and
important field in mathematical physics, yielding important
insights and results in diverse dimensions. One aspect of this
which has been of interest in recent years is the investigation of
topological gauge theories on manifolds of special holonomy. For
example, in four dimensions there is the Vafa-Witten theory,
arising from $N=4$, $d=4$ super Yang-Mills on manifolds with
$SU(2)$ holonomy [\VafaWitten]. In higher dimensions one also
finds topological gauge theories when one considers Yang-Mills
theories on manifolds of special holonomy -- in eight dimensions
there is a topological gauge theory defined on manifolds with
$Spin(7)$ holonomy and there are analogous theories on
seven-dimensional $G_2$ holonomy manifolds and Calabi-Yau three-
and four-folds [\VariousDim]. These theories have a rather
intriguing structure. What is more, they arise naturally in the
study of wrapped Euclidean branes in string theory, providing a
rich source of new insights and approaches.

It is natural to enquire if a similar set of topological gravity
theories exists. This is a little harder to answer, as
supergravity theories are rather more complicated. The first
topological gravity, in four dimensions, was found by Witten
[\WittenWeyl]. This is based on the Lagrangian which is the square
of the Weyl tensor, and the theory is defined on any Riemannian
four-manifold. Topological gravities on special holonomy manifolds
have yet to be fully described however. A natural theory to
elucidate first is the four-dimensional topological Einstein-Maxwell
gravity. This should arise from
$N=2$ supergravity on manifolds with $SU(2)$ holonomy, using the covariantly
constant spinors to twist the fermions. An earlier pioneering
analysis of this theory was
given in [\Fre]. A number of other papers also explored this
subject [\Laters]. Recently [\Baul], the geometric formulation of
this theory was clarified, and the action given up to quadratic
order in the fermions.

In this paper we present the {\it complete}
on-shell formulation of this four-dimensional topological gravity, with
the full local BRST symmetries. A new $SU(2)$ global R-symmetry is given and
we explain how the local Lorentz group is twisted to have a non-standard
action on the fields. We also give the twisted current multiplet
and its local BRST cohomology and show explicitly that the stress tensor
is exact under the local BRST symmetry. We further propose a new set of observables
for $SU(2)$ holonomy manifolds. All of these results are new.

As well as these new results, our work clarifies a number of issues
which have appeared opaque in previous works.
The discussion of twisting in [\Fre] involved an identification of
the local Lorentz and global R-symmetry group which is clearly problematic.
Our explicit use of Killing vectors on $SU(2)$ holonomy four-folds
makes the correct procedure quite clear, and further leads to a
theory with an explicit twisted {\it local}\  Lorentz symmetry, which
also makes it evident that the theory being considered has the
required diffeomorphism invariance.
In addition, since our BRST symmetry is also local,
our expressions are explicitly supercovariant, which is not
obvious in some earlier works such as [\Fre]. Also, the twisted BRSTs
in this reference do not square to zero on-shell as claimed,
since it is not consistent to simply drop higher order terms from the transformations
- one must write the full
transformations first, calculate their square, and only then
consider terms order
order in fermions, as we do in this paper.

The paper is organised as follows.
In section 2 we consider Euclidean $N=2$, $d=4$ supergravity
theory on an $SU(2)$ holonomy manifold. In section 3 we study the
resulting topological gravity theory, giving the twisted local BRST
symmetries and twisted local Lorentz symmetries. New residual global
R-symmetries of the twisted theory are also given. We show that both
the twisted action and the vierbein stress tensor are BRST-exact
and we give a set of BRST invariant topological operators derived
from descent equations. The complete twisted current multiplet and its
cohomology are also found. In section 4 we give the other twisted
supersymmetries of the theory, and we finish in section 5 with a
brief discussion of the context of these results and further
applications.


\chapter{Twisted $N=2$ four-dimensional supergravity and manifolds with
$SU(2)$ holonomy}

We begin with Euclidean $N=2$, $d=4$ supergravity [\FvN]. This has
as bosonic fields the vierbein $e_\mu^a$, and the graviphoton
$A_\mu$. The fermionic fields are two symplectic Majorana
gravitini which may be collected into one Dirac gravitino
$\chi_\mu$. World indices $\mu, \nu,...$ and local Lorentz indices
$a,b,...$ run from one to four. The Euclidean theory is invariant
under the bosonic symmetry $Spin(4) \times SU(2) \times
Spin(1,1)$ where $Spin(4) \cong {SU(2)}_+ \times {SU(2)}_-$ is the
local Lorentz group and $SU(2) \times Spin(1,1)$ is the global
R-symmetry group.
\foot{This is contrasted with the Lorentzian theory in $3+1$
dimensions which is invariant under $Spin(3,1) \times U(2)$. Both
Euclidean and Lorentzian theories arise from the reduction of the
Lorentzian $N=2$ supergravity in $5+1$ dimensions. This theory
[\NisSez] is invariant under $Spin(5,1) \times SU(2)$ and has 8
supersymmetries. The reduced Euclidean and Lorentzian theories
both preserve all 8 supersymmetries in four dimensions though the
Euclideanised theory obtained by Wick rotating the time component
in the Lorentzian theory preserves no supersymmetries. This clarifies [\Fre] what is meant by the initial "Euclidean" supergravity.}
Under the $Spin(1,1)$ R-symmetry subgroup,
the bosonic fields $e^a_\mu$ and $A_\mu$
 transform with weight zero and the fermionic
 fields $\chi_\mu$ (${\chi^{\dagger}_{\mu}}$) with weight $1/2$ ($-1/2$).

The supersymmetric action is given by [\TvN]
$$
\eqalign{ S^{N=2} = \int\! d^4x\ e\bigg(
 {1\over 2\kappa^2} &R(e,\hat\omega) + \epsilon^{\mu\nu\rho\sigma}
               \chi^\dagger_\mu\gamma_\nu {\hat{D}}_\rho \chi_\sigma
-{1\over4}F^2_{\mu\nu}  \cr
        & +{i\kappa\over2\sqrt2}\chi^\dagger_\mu\Big[ (F^{\mu\nu}+ \hat
F^{\mu\nu})
     + \gamma_5 \left( *F^{\mu\nu} + *\hat F^{\mu\nu} \right) \Big]\chi_\nu\bigg). \cr}
\eqn\Neqtwo
$$
Our spinor conventions are those of [\TvN]. $\kappa$ is the
gravitational constant and we define $*F_{\mu\nu} :=
{1\over2}\epsilon_{\mu\nu\rho\sigma}F^{\rho\sigma}$, with
$\epsilon_{\mu\nu\rho\sigma} = e^a_\mu e^b_\nu e^c_\rho e^d_\sigma
\epsilon_{abcd}$, and $\epsilon_{abcd}$ numerical.
The two-form
graviphoton field strength $F =dA$. $D:= D(\omega)$ is the
gravitational covariant derivative with respect to the standard
spin connection one-form $\omega^{ab}$, given by the solution of
the equation $De^a \equiv de^a + \omega^{ab} \wedge e^b = 0$. The
action of this covariant derivative on a spinor $\lambda$ is given
by $D\lambda \equiv d\lambda +
{1\over4}\omega^{ab}\gamma^{ab}\lambda$, where $\gamma_{ab}
:=\gamma_{[a}\gamma_{b]}$. The action of the supercovariant
derivative $\hat{D} := D(\hat{\omega} )$ on Lorentz tensors takes
the same form as that of $D$, but in terms of the supercovariant
spin connection one-form ${\hat{\omega}}^{ab}$ defined as the
solution of the equation ${\hat{D}}e^a \equiv de^a +
{\hat{\omega}}^{ab} \wedge e^b = - {\kappa^2 \over 2} \chi^\dagger
\wedge \gamma_5 \gamma^a \chi$. One can similarly write a
supercovariant graviphoton field strength ${\hat{F}}$
 which is defined below, together with the explicit expression for
 ${\hat{\omega}}^{ab}$:
$$
\eqalign{
\hat F_{\mu\nu}  & = F_{\mu\nu} - i\sqrt2\kappa
\chi^\dagger_{[\mu}\gamma_5\chi_{\nu]},  \cr
 \hat\omega^{ab}_\mu &= \omega^{ab}_\mu -{1\over4}\kappa^2 \Big(
\chi^\dagger_a\gamma_5\gamma_\mu\chi_b
        +\chi^\dagger_a\gamma_5\gamma_b\chi_\mu -
\chi^\dagger_\mu\gamma_5\gamma_b\chi_a
             - (a \leftrightarrow b) \Big). \cr}
\eqn\supercov
$$

The action \Neqtwo\  is invariant under the standard general coordinate,
local Lorentz and Maxwell
transformations, as well as the local supersymmetry
$$
\eqalign{
\delta_{\epsilon} e^a_\mu & = -\kappa\big( \epsilon^\dagger\gamma_5\gamma^a\chi_\mu -
                    \chi_\mu^\dagger\gamma_5\gamma^a\epsilon\big) , \cr
\delta_{\epsilon} A_\mu &=
i\sqrt2\big(\epsilon^\dagger\gamma_5\chi_\mu -
\chi_\mu^\dagger\gamma_5\epsilon\big), \cr \delta_{\epsilon}
\chi_\mu &= {2\over\kappa}\hat D_\mu\epsilon - {i\over\sqrt2}
                       \Big(\hat F_{\mu\lambda} \gamma^\lambda +
       *\!\hat F_{\mu\lambda} \gamma^\lambda \gamma_5\Big)\epsilon,
 \cr}
\eqn\susies
$$
where $\epsilon$ is a Dirac spinor.
We define chiral spinors by $\chi_L := {1\over2}(1+\gamma_5)\chi$,
$\chi_R := {1\over2}(1-\gamma_5)\chi$, with ${( \gamma_5 )}^2=1$.

On four-manifolds with $SU(2)$ holonomy, there exist two
covariantly constant chiral spinors which can be collected into a
complex spinor $\theta$ which we will take to be
right-handed.
\foot{This convention implies that the holonomy of the spin
connection is contained in the ${SU(2)}_-$ Lorentz subgroup, that is
generated by the anti-self-dual Lorentz generators.}
We will normalise this so that $\theta^\dagger\theta={1\over2}$.
On such manifolds, one may expand the gravitino one-form
$\chi$ in terms of the twisted anticommuting one-form fields
$\eta, \psi^a, \chi^{ab}$ as
$$
\eqalign{
\chi_R &= \eta\theta + {1\over2}\chi^{ab}\gamma_{ab}\theta, \cr
\chi_L &=  \psi^{a}\gamma_{a}\theta. \cr}
\eqn\twistedfields
$$
The field $\chi^{ab}$ is self-dual (satisfying $\chi^{ab} =
{1\over2}\epsilon^{abcd}\chi^{cd}$)
due to the chirality of $\theta$. We take the fields $\eta, \chi^{ab}$ to be
Hermitian, and $\psi^a$ to be anti-Hermitian.
One may similarly expand the local supersymmetry spinor parameter $\epsilon$ as
$$
\eqalign{
   \epsilon_R &  = \epsilon\theta  + {1\over2}\epsilon^{ab}\gamma_{ab}\theta,
\cr
     \epsilon_L &=  \epsilon^{a}\gamma_{a}\theta, \cr}
\eqn\twistedsusys
$$
where the $\epsilon, \epsilon^a, \epsilon^{ab}$ on the right-hand
sides of the equations above are anticommuting fields, with
$\epsilon^{ab}$ self-dual, and $\epsilon, \epsilon^{ab}$
Hermitian, and $\epsilon^a$ anti-Hermitian. These fields will
parametrise twisted supersymmetries in the twisted theory. We will
only consider the twisted scalar, or local topological BRST,
supersymmetry in the following section, turning to the other
twisted supersymmetries in section 4. Whilst the general
coordinate and Maxwell symmetries do not change under the
twisting, the local Lorentz symmetries of the action above,
generated as usual by ${1\over2}\gamma^{ab}$ on spinors, are
twisted on $SU(2)$ holonomy manifolds, resulting in a non-standard
action on the twisted fermion fields as we will see later.

With the above expansions of the gravitino and supersymmetry
parameter, one can now derive the twisted theory on a manifold
with $SU(2)$ holonomy. The resulting action is
$$
\eqalign{ S = \int \!d^4x &\ e\bigg( -{1\over\kappa^2}
\epsilon^{\mu\nu\rho\sigma}\omega_{\mu}^{+ab}e_\nu^b \,
\omega_{\rho}^{+ac}e_\sigma^c -
{1\over2}\epsilon^{\mu\nu\rho\sigma}(
D_{\mu}^{\scriptstyle{[1/2]}} \psi_\nu^a)(\eta_\rho e_\sigma^a -
    2\chi_\rho^{ab}e_\sigma^b) \cr
   & - {1\over2} F^-_{\mu\nu} \, F^{-\mu\nu}
   -{1\over2}\epsilon^{\mu\nu\rho\sigma}\psi_\mu^a\Big(( D_{\nu}^{\scriptstyle{[1/2]}} \eta_\rho)
   e_\sigma^a -
                    2( D_{\nu}^{\scriptstyle{[1/2]}} \chi_\rho^{ab})e_\sigma^b\Big) \cr
   & -{i\kappa\over2\sqrt2} ( F^{-\mu\nu} +
   {\hat{F}}^{-\mu\nu} )(\eta_\mu\eta_\nu+ \chi_\mu^{ab}\chi_\nu^{ab}) +
   {i\kappa\over2\sqrt2} ( F^{+\mu\nu} + {\hat{F}}^{+\mu\nu})(\psi^a_\mu\psi_\nu^a) \bigg) \cr}
\eqn\actionfirst
$$
where $\omega^{\pm ab}_\mu := {1\over2}( \omega^{ab}_\mu \pm
{1\over2} \epsilon^{abcd} \omega^{cd}_\mu )$, $F^\pm_{\mu\nu} :=
{1\over2}(F_{\mu\nu} \pm *F_{\mu\nu})$ and $D^{\scriptstyle [1/2]} :=
D({{\omega + \hat{\omega}}\over{2}})$. The pure
gravitational and Maxwell kinetic terms are unaffected by the
twist but have been rewritten using the identities
$$
\eqalign{
 {e\over2} R(e,\omega ) & \equiv - e\ \epsilon^{\mu\nu\rho\sigma} \omega^{+ab}_{\mu} e^{b}_{\nu}
 \omega^{+ac}_{\rho} e^{c}_{\sigma} + \partial_{\mu} \left(e\
  \epsilon^{\mu\nu\rho\sigma} \omega^{+ab}_{\nu} e^{a}_{\rho} e^{b}_{\sigma} \right), \cr
 {e\over4} F_{\mu\nu} F^{\mu\nu} & \equiv {e\over2} F^{-}_{\mu\nu}
 F^{-\mu\nu} + \partial_{\mu} \left( {e\over4} \epsilon^{\mu\nu\rho\sigma}
 A_{\nu} F_{\rho\sigma} \right), \cr }
\eqn\idents
$$
where $R^{ab} := R^{ab} (\omega) = d \omega^{ab} + \omega^{ac}
\wedge \omega^{cb}$ is the two-form Riemann tensor. The total
derivative terms in {\idents} have been discarded in the integral
{\actionfirst}.

Twisting the supersymmetry leads to the local BRST symmetry
$$
\eqalign{
\delta_{\epsilon} e_\mu^a &= \kappa\epsilon\psi_\mu^a, \cr
\delta_{\epsilon} A_\mu &= -i\sqrt2\epsilon\eta_\mu, \cr
\delta_{\epsilon} \eta_\mu &= {2\over\kappa}\partial_\mu\epsilon, \cr
\delta_{\epsilon} \psi_\mu^a &= -i\sqrt2\epsilon \hat F^+_{\mu\nu}e_a^\nu, \cr
\delta_{\epsilon} \chi_{\mu}^{ab} &= {1\over\kappa}\epsilon\, {\hat{\omega}}_{\mu}^{+ab}.
\cr }
\eqn\brstsfirst
$$

We will study this twisted theory
 in the following section, where we will show that it defines a
topological gravity theory of cohomological type.


\chapter{Topological four-dimensional gravity}

In the previous section we showed how Euclidean $N=2$, $d=4$
supergravity on an $SU(2)$ holonomy manifold becomes twisted,
giving a theory with anti-commuting one-form fields $\eta$,
$\psi^a$ and $\chi^{ab}$ (a self-dual Lorentz two-form), together
with the commuting one-form fields $e^a$ and $A$. In a similar
manner, the supersymmetry generator of the untwisted theory, which
is a Dirac spinor, reduces to a set of zero-form fermionic
generators $(Q, Q^a , Q^{ab} )$ (where $Q^{ab}$ is a self-dual
Lorentz two-form). We will consider here the scalar BRST-like
supersymmetry $Q$. The bosonic symmetries consist of the standard
general coordinate and Maxwell gauge transformations which remain
unchanged under the twist together with the local Lorentz
transformations whose action on the anti-commuting fields becomes
modified in the twisted theory, and further one has a rigid
$SU(2)$ R-symmetry, and an Abelian symmetry of the equations of
motion, as we will see.

It will prove convenient now to set $\kappa=2$ and to redefine the fields
 by $e^a \rightarrow i\sqrt2 e^a$, $A\rightarrow -i\sqrt2 A$, $\psi^a
  \rightarrow {i\over\sqrt2} \psi^a$. It is also convenient to introduce
  the second rank Lorentz tensor-valued one-form quantity $T^{ab} := \delta^{ab} \, \eta - 2\, \chi^{ab}$.
\foot{Since each component of $T^{ab}$ is in an irreducible representation
 of $SO(4)$ then one can reobtain the fields $\eta$ and $\chi^{ab}$ by Young projection on $T^{ab}$.}
Written in terms of differential forms, the action \actionfirst\ and BRST
 symmetry \brstsfirst\ then become
 (we drop an overall factor coming from the scaling of the integration measure under
 the above field redefinitions)
$$
\eqalign{S = \int &\bigg( {1\over2} (\omega^{+ab} \wedge e^b)
\wedge (\omega^{+ac} \wedge e^c) - 2 F^- \wedge F^- \cr
    &-{1\over2} T^{ab} \wedge e^b \wedge D^{\scriptstyle [1/2]}\psi^a
    +{1\over2} \psi^a \wedge e^b \wedge D^{\scriptstyle [1/2]} T^{ab} \cr
    &+{1\over4}{(F + {\hat{F}})}^- \wedge T^{ab} \wedge T^{ab} +
    {1\over2}{(F + {\hat{F}})}^+ \wedge \psi^a \wedge \psi^a \bigg)
 \cr}
\eqn\action
$$
and
$$
\eqalign{ \delta_{\epsilon} e^a &= \epsilon\psi^a, \cr
\delta_{\epsilon} A &= \epsilon\eta, \cr
\delta_{\epsilon} \eta &=d\epsilon, \cr
\delta_{\epsilon} \psi^a &= -2\epsilon{\hat{F}}^{-ab} e^b, \cr
\delta_{\epsilon} \chi^{ab} &={1\over2}\epsilon\, {\hat{\omega}}^{+ab}. \cr}
\eqn\brsts
$$
World and Lorentz indices are related using vierbein
$e^{a}_{\mu}$, with $F_{ab} = F_{\mu\nu}e_a^\mu e_b^\nu$. The
differential form notation is introduced here for simplicity.

The local Lorentz symmetries are given by twisting
the original local Lorentz transformations, giving
$$
\eqalign{
\delta_{L} e^a &= \Lambda^{ab}e^b, \cr
\delta_{L} A &= 0, \cr
\delta_{L} \eta &= -{1\over2}\Lambda^{+ab}\chi^{ab}, \cr
\delta_{L} \psi^a &= \Lambda^{-ab}\psi^b, \cr
\delta_{L} \chi^{ab} &= {1\over2} \Lambda^{+ab}\eta +
\Lambda^{+c[a}\chi^{b]c},\cr
}
\eqn\LLs
$$
for antisymmetric local Lorentz parameter $\Lambda^{ab} (x)$. The composite field $T^{ab}$
transforms nicely under these twisted Lorentz transformations with $\delta_{L} T^{ab} =
\Lambda^{+bc} T^{ac}$. Notice how the self-dual part of
the local Lorentz parameter $\Lambda^{ab}$ acts only on
$\eta$ and $\chi^{ab}$, whilst $\psi^{a}$ transforms only
under the anti-self-dual part (as expected, since the action
is localised on self-dual connections and $\psi^a$ corresponds
to vierbein variations preserving this condition [\Baul]).

The associated covariant derivatives $D$ can be written schematically as $D[-] =
d[-] + \delta_{L} (\omega) \wedge [-]$, so that
$$
\eqalign{
D\eta &= d\eta - {1\over2} \omega^{+ab} \wedge \chi^{ab}, \cr
D\psi^a &= d\psi^a + \omega^{-ab} \wedge \psi^b,\cr
D\chi^{ab} &= d\chi^{ab} + {1\over2} \omega^{+ab} \wedge \eta +
\omega^{+c[a} \wedge \chi^{b]c}.\cr
}
\eqn\CovDerivs
$$

The action \action\ has a further rigid $SU(2)$ R-symmetry, given
by
$$
\eqalign{
 \delta_{\lambda} e^a &= \lambda^{+ab}e^b, \cr
  \delta_{\lambda} A &= 0,\cr
  \delta_{\lambda} \eta &= 0, \cr
\delta_{\lambda} \psi^a &= \lambda^{+ab}\psi^b, \cr
 \delta_{\lambda} \chi^{ab}&=\lambda^{+c[a}\chi^{b]c},\cr }
 \eqn\Rsymmetry
$$
where $\lambda^{+ab}$ is the constant self-dual parameter of the
symmetry. This symmetry is independent from the local Lorentz invariance under {\LLs}
with global parameter $\Lambda^{ab}(x) = \Lambda^{ab}$. Note however that the transformations
corresponding to {\Rsymmetry} with a constant anti-self-dual parameter $\lambda^{-ab}$ are a
special case of the local Lorentz transformations, with $\Lambda^{ab} = \lambda^{-ab}$.

The supercovariant and local Lorentz invariant graviphoton field
strength ${\hat{F}}$ in {\action}, {\brsts} is given by
$$
{\hat{F}} = F - {1\over8} T^{ab} \wedge T^{ab} + {1\over4} \psi^a
\wedge \psi^a.
$$
The supercovariant spin connection $\hat\omega(e,T,\psi)$ is the
solution of the equation ${\hat{D}} e^a \equiv  {( {\hat{\omega}}
- \omega )}^{ab} \wedge e^b = T^{ba} \wedge \psi^b$. The explicit
solution is given by
$$
{\hat{\omega}}_{\mu}^{ab} = \omega_{\mu}^{ab} - 2 e^{\nu [a} T^{|c|b]}_{[\mu }
 \psi^{c}_{\nu ]} + e^{\rho a} e^{\sigma b} e^{c}_{\mu} T^{dc}_{[\rho } \psi^{d}_{\sigma ]}.
$$

The supercovariant objects just defined follow from twisting the corresponding
objects in the previous section.
We note in passing the result
$$
(D^{\scriptstyle [1/2]}\psi^a) \wedge T^{ab} \wedge e^b
=
\psi^a \wedge ( D^{\scriptstyle [1/2]} T^{ab} ) e^b - {1\over{16}} T^{ab}
\wedge T^{ab} \wedge \psi^c \wedge \psi^c,
$$
up to a total derivative, implying that terms quartic in the
fermions are generated in the twisted
theory if one writes the action with only left-
or right- acting covariant derivatives in the fermion kinetic terms.

The explicit BRST-supercovariant form of the twisted action and
  symmetries above clarifies the results in [\Fre] and the
   non-standard geometrical symmetries {\LLs}, {\Rsymmetry} are new.

It should be noted that once having used the $SU(2)$ structure
to twist, the resulting action {\action} is invariant under all
 the transformations above on $any$ Riemannian four-manifold.
 Writing this twisted theory on a general four-manifold however
  necessitates a non-standard action of the local Lorentz group
  on fields {\LLs}, {\CovDerivs} in order to maintain covariance.

We will now show that the model detailed above satisfies the
axioms
 of a topological field theory of cohomological type. That is, the action
 of the twisted scalar BRST operator $Q$ forms an on-shell, (Lorentz + gauge)
  equivariant cohomology in which the action functional is closed and the
   associated stress tensor is exact. It will
    also be shown that the action functional itself is BRST-exact.

\section{BRST invariance of the action}

This follows from the supersymmetry of the untwisted theory,
nevertheless an explicit calculation is a useful check of the
result. The following procedure simplifies the calculation.
Writing the set of one-form fields in the twisted theory as $\{
\Phi \}$ then it is evident that under an arbitrary infinitesimal
field transformation $\delta \Phi$, the action $S$ varies as
$$
\delta S = \int \! \sum_{\Phi} ( \delta \Phi ) \wedge J_{\Phi}
$$
where all total derivative terms are dropped. $\{ J_{\Phi} \}$ form a set of
three-forms associated with $\{ \Phi \}$. Each of these three-forms are Hodge-dual
to the equations of motion for the respective fields in the sense that
$[ \Phi ]:= {{\delta S}\over{\delta \Phi}} = * J_{\Phi} =0$ is the field equation for $\Phi$.

To lowest order in the fermionic fields, it is found that under the local BRST
transformations {\brsts}, the $[ e^a ]$ terms in the variation of the twisted
action {\action} cancel all the $[ T^{ab} ]$ terms and also various $[ \psi^a ]$
terms. The $[ A ]$ terms cancel the remaining $[ \psi^{a} ]$ terms. The action
{\action} is also found to be invariant under {\brsts} to all orders with various
 additional cross-cancellations occuring.

The fact that this calculation holds for local BRST transformations can be seen
 from the fact that the three-form field equation tensors
$$
\eqalign{
J_{e^a} &= - e^b \wedge R^{ab} ( {\hat{\omega}}^+ ) - \psi^b \wedge {\hat{D}}
T^{ba} -4 {\hat{F}}^{-ab} e^b \wedge {\hat{F}}^+ , \cr
J_{A} &= -d \left( 4 {\hat{F}}^- - \psi^a \wedge \psi^a \right) , \cr
J_{T^{ab}} &= - e^b \wedge {\hat{D}} \psi^a + T^{ab} \wedge {\hat{F}}^- , \cr
J_{\psi^a} &= e^b \wedge {\hat{D}} T^{ab} + 2 \psi^a \wedge {\hat{F}}^+ , \cr }
\eqn\eqnsmotion
$$
are explicitly BRST-supercovariant.

The equations of motion \eqnsmotion\ are invariant under
the transformations
$$
\eqalign{
 \delta_{\alpha} e^a &= 0, \cr
  \delta_{\alpha} \hat F^- &= -2 \alpha \hat F^-,\cr
 \delta_{\alpha} \hat F^+ &= 2 \alpha \hat F^+,\cr
\delta_{\alpha} \eta &= \alpha\eta, \cr
 \delta_{\alpha} \psi^a &=- \alpha\psi^a, \cr
 \delta_{\alpha} \chi^{ab}&= \alpha\chi^{ab},\cr }
 \eqn\eqnmsymmetry
$$
for real parameter $\alpha$. These BRST-supercovariant transformations define the
(non-local in $A$) transformations of $F^{\pm}$.

Some identities used above and in various calculations throughout this paper are
$$
\eqalign{
& A^{-}_{ac} B^{+}_{bc} = A^{-}_{bc} B^{+}_{ac}, \cr
& A^{\pm}_{ac} B^{\pm}_{bc} + A^{\pm}_{bc} B^{\pm}_{ac} =
{1\over2} \delta_{ab} A^{\pm}_{cd} B^{\pm}_{cd}, \cr
& A^{\pm}_{ae} B^{\pm}_{be} - A^{\pm}_{be} B^{\pm}_{ae} \mp \epsilon_{abcd}
A^{\pm}_{ce} B^{\pm}_{de} = 0,  \cr}
$$
for any antisymmetric tensors $A_{ab}$, $B_{ab}$.

Note also that the tensor $\epsilon_{\mu\nu\rho\sigma}$ depends on the vierbein. 
This results in additional variations of the vierbein when one varies the
(anti-)self-dual parts $A^{\pm}_{\mu\nu}$ of a two-form $A_{\mu\nu}$.
If the variation of $A$ is $\delta A$ then
$$
\delta \left( A^{\pm} \right) = {\left( \delta A \right)}^{\pm} -
{( \delta e^a \wedge e^b )}^{\pm} A^{\mp}_{ab},
$$
where $A_{ab} = A_{\mu\nu} e^{\mu}_{a} e^{\nu}_{b}$.

\section{Nilpotence of the BRST symmetry}

The local BRST transformations \brsts\ are equivariantly nilpotent
on-shell, satisfying
$[\delta_{\epsilon},\delta_{\epsilon^{\prime}}]=0$ up to fermionic
equations of motion, local Lorentz transformations and $U(1)$
gauge transformations. The local Lorentz transformations have
parameter $\Lambda^{ab} = 4\epsilon \epsilon^{\prime}
{\hat{F}}^{-ab}$, whilst the gauge transformations have parameter
$-\epsilon \epsilon^{\prime}$. The proof of BRST nilpotence
requires showing that $\delta_{\epsilon} \, {\hat{\omega}}^{+ab}
=0$ and $( \delta_{\epsilon} {\hat{F}}^{-ab} ) e^b=0$ on-shell. It
is straightforward but rather tedious to show that these equations
are true. This involves taking various contractions of the $\eta$,
$\psi^a$ and $\chi^{ab}$ equations of motion with the inverse
vierbein. For example, taking contractions of the vierbien with
the $\psi^a$ equation of motion, one deduces that
$$
\delta_{\epsilon} \Big( F_{\mu\nu} - \eta_\mu \eta_\nu - \chi^{ab}_{\mu} \chi^{ab}_{\nu} \Big)
\vert_{[\mu\nu]-} =
         2 \epsilon \Big( {\hat{D}}_{[\mu} \eta_{\nu ]} + e^a_{(\mu}\psi^a_{\lambda)}
                           {\hat{F}}^{+\lambda}_\nu \Big)\vert_{[\mu\nu]-} = 0,
$$
on-shell. This result is used to prove that the BRST symmetry is nilpotent when acting
on $\psi^a$.
More complicated manipulations of contractions of the vierbein with the
$\eta$ and
$\chi^{ab}$ equations of motion yield the result that
$$
\Big(e^{\nu a}\hat D_{[\nu}\psi^b_{\mu]} + \hat F^-_{\mu\nu}
e^{\nu a} e^{\rho b}\eta_\rho
     -2 \hat F^-_{\mu\nu} e^{\nu c}
e^{\rho b}\chi_\rho^{ca}\Big)\vert_{[ab]} = 0,
$$
on-shell. This result is used to prove that the BRST symmetry is nilpotent when acting on $\chi^{ab}$. Note that this calculation only works by using the full supercovariant BRST rules {\brsts}. The lowest order on-shell global BRST transformations given in (4.3) of [\Fre] do not square to zero (even to lowest order on-shell and up to local transformations).

\section{Exactness of the action}

Not only is the action {\action} BRST-closed, it is also
BRST-exact, up to the fermion equations of motion. This is in the
sense that if one writes the action $S = \int d^4x L$ then one has
$\int d^4 x\, \epsilon L = \delta_{\epsilon} \Psi$, where the
gauge fermion $\Psi$ is given by
$$
\eqalign{ \Psi = \int \! &\left( \chi^{ab} \wedge e^b \wedge
\left( {\hat{\omega}}^{ac} \wedge e^c - 2\eta \wedge \psi^a
\right) \right. \cr &\left.+ {1\over2} e^a \wedge \psi^a \wedge
\Big( {\hat{F}}^- + {1\over4} T^{bc} \wedge T^{bc} - {1\over4}
\psi^b \wedge \psi^b \Big) \right). \cr} \eqn\gaugefermion
$$
This is fairly straightforward to prove if one notes that since one is
allowed to use the
fermion equations of motion, one can use these to generate the terms in the
action which are
quadratic in the fermions. This will also generate other terms quartic in the
fermions.
Combining these with the results of performing the BRST variation of the
gauge fermion
above, one proves that precisely  the action \action\ is generated. Note also
that one assumes that the BRST parameter $\epsilon$ is constant for this calculation.
Any terms
involving $d \epsilon$ can be absorbed as total derivatives into
the Lagrangian.

It should be noted that the gauge fermion above is invariant under
the $\Lambda^{-ab}$ part of the local Lorentz transformation
$\delta_{L} ( \Lambda^{ab} )$, but not under the $\Lambda^{+ab}$
part. This is consistent with localisation in the path integral.
That is, one can introduce a coupling constant in front of the
Lagrangian, and use exactness of the action to show that
correlation functions are independent of the value of this
coupling constant. This can be used to prove that the path
integral localises on field configurations with $\omega^{+ab}
\wedge e^b =0$ and $F^- =0$. The former is only invariant under
the $\Lambda^{-ab}$ local Lorentz transformations.

In the gravitational sector, localisation on vierbeins satisfying
the anti-self-dual spin connection constraint $\omega^{+ab}_{\mu}
=0$ implies localisation on metrics satisfying the constraint
$R^{ab} + {1 \over 2} \epsilon^{abcd} R^{cd} =0$, due to the
identity $R^{ab} (\omega^+ ) = R^{+ab} ( \omega )$. Gravitational
instanton metric solutions of this equation have been well studied. Note that
under the standard Kaluza-Klein reduction of the four-dimensional metric
$g_{\mu\nu} \rightarrow ( e^{\phi / \sqrt{3}} g_{ij} + e^{-2 \phi / \sqrt{3}} A_i A_j ,
   e^{-2 \phi / \sqrt{3}} A_i , e^{-2 \phi / \sqrt{3}} )$ to three dimensions,
this anti-self-duality condition reduces to two independent, covariant constraints. 
The first just demands that one considers flat spin connections in three dimensions 
whilst the second imposes the non-linear Bogomolnyi type equation
$$
F_{ij} + e^{\phi} \epsilon_{ijk} \partial^k \phi =0,
\eqn\bog
$$
where $F_{ij} = 2 \partial_{[i} A_{j ]}$ and $\epsilon_{ijk}$ denotes
the orientation of the three-manifold. This is analogous to the 
reduction of Donaldson-Witten theory where one obtains Floer 
theory in three dimensions whose path integral localises on 
solutions of the standard Bogomolnyi equation. The topological 
gravity in three dimensions which localises on solutions of {\bog} would 
analogously be obtained by simple Kaluza-Klein reduction of the theory presented here.

\section{Exactness of the stress tensor}

The action \action\ is a functional of the vierbien $e^a_\mu$, explicitly as
well as through
its dependence on the metric $g_{\mu\nu} = e^a_\mu e^a_\nu$. Variation of the action with respect
 to the (inverse) vierbein generates the \lq\lq vierbein stress tensor"
\foot{This is related to the usual stress tensor
$T_{\mu\nu} = {2\over\sqrt g}{\delta S\over\delta g^{\mu\nu}}$
by $
T_{\mu\nu}  = e^a_\nu V^a_\mu$.
Conservation and symmetry of the stress tensor follow from the general
coordinate
and local Lorentz invariance of the action (see [\Weinberg]).}
$$
V^a_\mu = {1\over e} {\delta S\over\delta e^\mu_a} \equiv {1\over e} [ e_{\mu}^a ].
$$
Thus the vierbein stress tensor is given (up to a factor of e)
 by the equation of motion for the
vierbein. The expression for $V^{a}$ (as a one-form) then simply follows from $[ e^a ]
= * J_{e^a}$ defined in {\eqnsmotion}.

Since the BRST transformations are a symmetry of the theory, they must
take equations of motion into other equations of motion. In particular,
a calculation shows that the BRST variation of the field equation three-forms {\eqnsmotion} give
$$
\eqalign{ \delta_{\epsilon} J_{\psi^a} &= \epsilon\, J_{e^a}, \cr
 \delta_{\epsilon} J_{\eta} &= \epsilon\, J_{A}, \cr
 \delta_{\epsilon} J_{\chi^{ab}} &= 0. \cr}
\eqn\brsteq
$$
These relations follow even for local BRST transformations since the field equations
{\eqnsmotion} are supercovariant. The implication of {\brsteq} is that
the stress tensor is BRST-exact, on-shell, such that
$$
\epsilon\, T_{\mu\nu}  = \delta_{\epsilon} G_{\mu\nu}, \eqn\stress
$$
for some function $G_{\mu\nu}$. This statement follows since
the vierbein stress tensor $V^{a}_{\mu} = T_{\mu\nu} e^{\nu a}
 = {1\over e} {( * J_{e^a} )}^{a}_{\mu}$. Therefore, using
 {\brsteq} and the fact the $* J_{\psi^a} =0$ is a fermionic
 equation of motion, one has the result {\stress}.
Thus the local BRST transformations provide the twisted theory
with a cohomological structure.

In the off-shell, gauge-fixed formulation of twisted gauge theories then exactness of the stress tensor follows from exactness of the action. This is why the calculation above is not done in [\Fre]. For twisted gravitational theories though, gauge-fixing implies fixing diffeomorphism symmetry and then it is not so clear whether the statement above is still valid. 

\section{Descent equations and invariants}
To construct a basis of BRST-invariant topological operators,
consider a set of operators of the form ${\cal{O}}_i =
\int_{\gamma_i} W_{(i)}$ (where $\gamma_i$ is an $i$-cycle in
homology and $W_{(i)}$ is an $i$-form). These operators are
constructed in the standard way, via the descent equations $Q\cdot
W_{(i+1)} = d W_{(i)}$.
\foot{We here consider rigid BRST transformations
$\delta_{\epsilon}$ {\brsts}, which
 can be factorised such that $\delta_{\epsilon} = \epsilon Q$, defining the action of $Q$ on the fields.}
These equations are solved by starting with a
four-form characteristic class $W_{(4)}$ whose
 evaluation on the four-manifold produces a
 topological invariant. Recall that for
 twisted gauge theories in four dimensions, this is given by the
 second Chern class of the gauge bundle which is proportional to $F \wedge F$ and
 whose evaluation gives the instanton number. For twisted gravity in four dimensions
 there are two natural four-form characteristic classes on the tangent bundle of the
  four-manifold. These are the first Pontrjagin class (proportional to $R^{ab} \wedge R^{ab}$)
  and the Euler class (proportional to $\epsilon^{abcd} R^{ab} \wedge R^{cd}$) whose evaluation
  gives the Hirzebruch signature and Euler number invariants, respectively. It is convenient to take linear
  combinations of these two classes to define $W^{\pm}_{(4)} := R^{\pm ab} \wedge R^{\pm ab}$.
  The set of descendent
  BRST-invariant operators can then be written
$$
\eqalign{
{\cal{O}}^{\pm}_{4} &= \int_{\gamma_4} R^{\pm ab} \wedge R^{\pm ab}, \cr
{\cal{O}}^{\pm}_{3} &= \int_{\gamma_3} 2 N^{\pm ab} \wedge R^{\pm ab}, \cr
{\cal{O}}^{+}_{2} &= \int_{\gamma_2} N^{+ab} \wedge N^{+ab}, \cr
{\cal{O}}^{-}_{2} &= \int_{\gamma_2} \left( 4 {\hat{F}}^{-ab} R^{-ab} +
N^{-ab} \wedge N^{-ab} \right), \cr
{\cal{O}}^{-}_{1} &= \int_{\gamma_1}  4 {\hat{F}}^{-ab} N^{-ab}, \cr
{\cal{O}}^{-}_{0} &= \int_{\gamma_0} 4 {\hat{F}}^{-ab} {\hat{F}}^{-ab}. \cr
 }
\eqn\observ
$$
The anticommuting Lorentz two-form valued one-form $N^{ab} := Q\cdot \omega^{ab}$ and
${\cal{O}}^{+}_{1} = {\cal{O}}^{+}_{0} =0$. Each operator in {\observ} is $Q$-invariant,
though for ${\cal{O}}^{\pm}_{2}$, ${\cal{O}}^{-}_{1}$ and ${\cal{O}}^{-}_{0}$ this requires
the fermionic field equation $( Q\cdot {\hat{F}}^{-ab} ) e^{b}_{\mu} =0$. These BRST supercovariant operators are related to those proposed in [\Fre] after identifying ${\hat{F}}^{ab}$ and $N^{ab}$ with their appropriate ghost fields, called ${\eta_0}^{ab}$ and ${\chi_0}^{ab}$ respectively in [\Fre].

There is also the Chern
class of the gauge bundle from which one can
 construct topological operators. The associated BRST-invariant operators are
$$
\eqalign{
{\cal{O}}^{\prime}_{4} &= \int_{\gamma_4} F \wedge F, \cr
{\cal{O}}^{\prime}_{3} &= \int_{\gamma_3} 2 \eta \wedge F, \cr
{\cal{O}}^{\prime}_{2} &= \int_{\gamma_2} \eta \wedge \eta, \cr
 }
\eqn\gaugeobserv
$$
which are $Q$-invariant off-shell.

We note that, up to boundary terms, the integrands in
${\cal{O}}^{\pm}_3$, ${\cal{O}}^{+}_2$ and ${\cal{O}}^{-}_1$ are $Q$-exact (for
${\cal{O}}^{-}_1$ this uses the fermionic field equation $( Q\cdot {\hat{F}}^{-ab} ) e^{b}_{\mu} =0$).
For example, the BRST variations of the Chern-Simons three-forms for $\omega^{\pm ab}$ yield the
integrands in ${\cal{O}}^{\pm}_3$ up to a total derivative.
The integrands in ${\cal{O}}^{\prime}_{2}$ and ${\cal{O}}^{\prime}_{3}$ are also $Q$-exact.
We remark that $SU(2)$ holonomy manifolds are necessarily Ricci-flat. Thus, for these manifolds, 
using {\idents}, 
$\omega^{+ab} \wedge e^b \wedge \omega^{+ac} \wedge e^c = d ( \omega^{+ab} \wedge e^a \wedge e^b )$,
so that locally this term in the Lagrangian is a total derivative. It is therefore potentially a four-form characteristic class on four-manifolds with holonomy in $SU(2)$ that would lead to a new set of descendent topological observables.

\chapter{Further twisted supersymmetries}

The topological gauge theory presented here has been shown to have local BRST invariance.
However, as can been seen from {\twistedsusys}, twisting the supersymmetry of
the original Euclidean $N=2$, $d=4$ supergravity would then be expected to yield further
 local fermionic symmetries.
This is the case, and one has a vector supersymmetry, with anti-commuting parameter
$\epsilon^a$, acting as
$$
\eqalign{
\delta_{\epsilon^a} e^a &= \epsilon^b T^{ba}, \cr
\delta_{\epsilon^a} A &= -{1\over2}\epsilon^a\psi^a, \cr
\delta_{\epsilon^a} \eta &= \epsilon^a\hat F^{+ab}e^b, \cr
\delta_{\epsilon^a} \chi^{ab} &= -2 {\hat{F}}^{+c[a} \epsilon^{b]+} e^c, \cr
\delta_{\epsilon^a} \psi^a &= d\epsilon^a + {\hat{\omega}}^{-ab} \epsilon^b \equiv
 {\hat{D}} \epsilon^a. \cr
 }
\eqn\vectorsusy
$$
There is also an antisymmetric tensor supersymmetry, with anti-commuting, self-dual
parameter $\epsilon^{ab}$, acting as
$$
\eqalign{
\delta_{\epsilon^{ab}} e^a &= 2\epsilon^{ab}\psi^b, \cr
\delta_{\epsilon^{ab}} A &= \epsilon^{ab}\chi^{ab}, \cr
\delta_{\epsilon^{ab}} \eta &= -{1\over2}\epsilon^{ab}\hat\omega^{+ab}, \cr
\delta_{\epsilon^{ab}} \chi^{ab} &= d\epsilon^{ab} +\hat\omega^{+c[a}\epsilon^{b]c}
 \equiv {\hat{D}} \epsilon^{ab} - {1\over2} \epsilon^{ab} \eta, \cr
\delta_{\epsilon^{ab}} \psi^a &= 4 \epsilon^{ab} \hat F^{-bc} e^c. \cr
 }
\eqn\vectorsusy
$$

The supersymmetry algebra of Euclidean $N=2$, $d=4$ supergravity
closes on-shell, modulo (field-dependent) diffeomorphisms, local
Lorentz transformations, gauge transformations and
supersymmetries. This is reflected in the brackets of the twisted
 supersymmetries $( \delta_{\epsilon} , \delta_{\epsilon^a} , \delta_{\epsilon^{ab}} )$
  and Lorentz transformations $\delta_{L}$, given by
$$
\eqalign{
[ \delta_{\epsilon} , \delta_{\epsilon^\prime} ] =&\, \delta_{L} (4\epsilon \epsilon^\prime
{\hat{F}}^{-ab}) + \delta_{\scriptstyle gauge} ( -\epsilon \epsilon^\prime ), \cr
[ \delta_{\epsilon^a} , \delta_{{\epsilon^\prime}^a} ] =&\, \delta_{L} (-2
\epsilon^c {\epsilon^\prime}^c {\hat{F}}^{+ab}) + \delta_{\scriptstyle gauge}
( {1\over2} \epsilon^a {\epsilon^\prime}^a ), \cr
[ \delta_{\epsilon^{ab}} , \delta_{{\epsilon^\prime}^{ab}} ] =&\, \delta_{L}
(4 \epsilon^{cd} {\epsilon^\prime}^{cd} {\hat{F}}^{-ab}) + \delta_{\scriptstyle
gauge} ( - \epsilon^{ab} {\epsilon^\prime}^{ab} ), \cr
[ \delta_{\epsilon} , \delta_{{\epsilon^\prime}^a} ] =&\, \delta_{L} (- \epsilon
 {\epsilon^\prime}^\mu {\hat{\omega}}_{\mu}^{ab}) + {\cal{L}}_{( -\epsilon
 {\epsilon^\prime}^\mu )} + \delta_{\scriptstyle gauge} ( \epsilon {\epsilon^\prime}^\mu A_\mu ) \cr
&+ \delta_{( \epsilon {\epsilon^\prime}^\mu \eta_\mu )} + \delta_{( \epsilon
{\epsilon^\prime}^\mu \chi^{ab}_\mu )} + \delta_{( \epsilon {\epsilon^\prime}^\mu \psi^a_\mu )}, \cr
[ \delta_{\epsilon} , \delta_{{\epsilon^\prime}^{ab}} ] =&\, 0, \cr
[ \delta_{\epsilon^a} , \delta_{{\epsilon^\prime}^{ab}} ] =&\, \delta_{L}
(2 \epsilon^c {\epsilon^\prime}^{c\mu} {\hat{\omega}}_{\mu}^{ab}) +
{\cal{L}}_{(2 \epsilon^a {\epsilon^\prime}^{a\mu} )} +
\delta_{\scriptstyle gauge} (-2 \epsilon^a {\epsilon^\prime}^{a\mu} A_\mu ) \cr
&+ \delta_{(-2 \epsilon^a {\epsilon^\prime}^{a\mu} \eta_\mu )} +
 \delta_{(-2 \epsilon^c {\epsilon^\prime}^{c\mu} \chi^{ab}_\mu )} +
 \delta_{(-2 \epsilon^b {\epsilon^\prime}^{b\mu} \psi^a_\mu )}, \cr
[ \delta_{\epsilon} , \delta_{L} ( \Lambda^{ab} ) ] =&\,
 \delta_{( {1\over2} \epsilon \Lambda^{+ab} )}, \cr
[ \delta_{\epsilon^a} , \delta_{L} ( \Lambda^{ab} ) ] =&\,
 \delta_{( \epsilon^b \Lambda^{-ab} )}, \cr
[ \delta_{\epsilon^{ab}} , \delta_{L} ( \Lambda^{ab} ) ] =&\,
\delta_{( -{1\over2} \epsilon^{ab} \Lambda^{+ab} )} +
\delta_{( - \epsilon^{c[a} \Lambda^{+b]c} )}. \cr
 }
\eqn\brackets
$$
The notation in the right hand side of {\brackets} is
 that ${\cal{L}}_{k^{\mu}}$ denotes the Lie derivative
 of a tensor under an infinitesimal general coordinate
 transformation $\delta x^{\mu} = - k^{\mu}$, whilst
 $\delta_{\scriptstyle gauge} ( \xi )$ denotes an
 infinitesimal $U(1)$ gauge transformation
 with parameter $\xi$. World and Lorentz indices
 are interchanged using the vierbein and its inverse.

Typically these other twisted supersymmetries are not preserved for twisted gauge
theories on a general curved background and are not given in [\Fre] though for
 twisted supergravities they are a well-defined symmetry.


\chapter{Discussion}

We have presented here the complete on-shell action and symmetries
for topological four-dimensional Einstein-Maxwell gravity,
together with a set of associated
topological operators. Topological gravities on higher-dimensional
special holonomy manifolds are expected to take a similar form to
the theory discussed here on manifolds of $SU(2)$ holonomy. This
was conjectured in [\Baul], and recently an eight-dimensional
$Spin(7)$ model has been analysed in [\BaulTwo]. We can thus
expect that all the special holonomy topological
gravities will be found soon. This will fill out the analogy with
Yang-Mills theories and their associated special holonomy
topological gauge theories. The special holonomy topological
Yang-Mills theories arise as the effective theories on Euclidean
branes wrapped on special holonomy manifolds. (More generally
these topological theories should be of Born-Infeld type, although
this formulation is only known for the Abelian Donaldson-Witten
theory [\bill]). The way in which these new topological gravities
might arise in string theory is an interesting issue. They may be
realised as low energy effective actions for twisted string
 theories, for example. This may occur in the context of
 duality, since there are examples or conjectured cases
 where topological gravities or string theories arise from
 holographic dualities. Three-dimensional Chern-Simons theory is
 known to be dual to a topological string theory (see [\vafaetal] and
 references therein). Also, a recently discovered topological conformal
 field theory in four dimensions is conjectured to be dual, via de Sitter holography,
 to a five-dimensional topological phase of type ${IIB}^*$, $d=9+1$ supergravity in the 
't Hooft limit [\dfhs]. The latter is expected to have, as boundary theory, a twisted 
version of four-dimensional $N=4$ Euclidean conformal supergravity.
 The special holonomy topological gravities, whilst of interest in
 themselves, may also find a natural setting in string theory.

\refout

\vfill\break

\bye